\documentstyle[psfig]{article}

\begin{document}

\title{Extraction of level density and $\gamma$ strength function from
primary $\gamma$ spectra}

\author{A. Schiller\footnote{Electronic address: Andreas.Schiller@fys.uio.no},
L. Bergholt, M. Guttormsen, E. Melby, J. Rekstad,\\ 
and S. Siem\\
Department of Physics, University of Oslo,\\ 
P.O.Box 1048, Blindern, N-0316 Oslo, Norway}

\date{}

\maketitle

\begin{abstract}
We present a new iterative procedure to extract the level density and the
$\gamma$ strength function from primary $\gamma$ spectra for energies close up
to the neutron binding energy. The procedure is tested on simulated spectra and
on data from the $^{173}$Yb($^3$He,$\alpha$)$^{172}$Yb reaction.

\noindent
PACS number(s): 29.85.+c, 21.10.Ma, 25.55.Hp, 27.70.+q
\end{abstract}

\section{Introduction}

The $\gamma$ transitions of excited nuclei give rich information on nuclear 
properties. In particular, the energy distribution of the first emitted 
$\gamma$ rays from a given excitation energy reveals information on the level 
density at the excitation energy to which the nucleus decays, and the $\gamma$ 
strength function at the difference of those two energies. If the initial and 
final excitation energy belong to the continuum energy region, typically above 
4~MeV of excitation energy for nuclei in the rare earth region, also 
thermodynamical properties may be investigated 
\cite{MB99,SB99}.

Recently, the nuclear level density has become the object of new interest. 
There is strong theoretical progress in making calculations applicable to 
higher energies and heavier nuclei. In particular, the shell model Monte Carlo 
technique \cite{LJ93,KD97} moves frontiers at present, and it is now mandatory 
to compare these calculations with experiments. Furthermore, the level density 
is essential for the understanding of the nucleon synthesis in stars. The level
densities are input in large computer codes where thousands of cross sections 
are estimated \cite{Go96}.

Our present knowledge of the gross properties of the $\gamma$ strength function
is also poor. The Weisskopf estimate which is based on single particle 
transitions, see e.g.~\cite{BM69}, gives a first estimation for the strengths. 
However, for some measured $\gamma$ transitions the transition rate may deviate
many orders of magnitude from these estimates. A recent compilation on average 
$\gamma$ transition strengths for M1, E1 and E2 transitions is given in
Ref.~\cite{ZS93}. The uncertainties concern the absolute strength as well as 
how the strength depends on the $\gamma$ transition energy. For E1 transitions,
it is usually assumed that the energy dependency follows the GDR 
$(\gamma,\gamma')$ cross section, however, this is not at all clear for low 
energy $\gamma$ rays.

In this work we describe a method to extract simultaneously the level density 
and $\gamma$ strength function in the continuum energy region for low spin
(0-6~$\hbar$). The basic ideas and the assumptions behind the method were first
presented in Ref.~\cite{HB95}. An implementation using an iterative projection 
technique, was first described in Ref.~\cite{TB96}. However, due to the 
existence of infinitely many solutions and the unfortunate renormalization of 
the primary $\gamma$ spectrum in every iteration step, this first 
implementation suffered from various severe problems, including divergence of 
the extracted quantities \cite{BG97}. Several solutions of the convergence 
problem have been proposed and presented at different conferences, using 
approximate normalizations, but none of them yielding exact reproductions of 
test spectra. However, data using one of those approximate methods were 
published in Ref.~\cite{MB99}. Today, we consider the previous methods as
premature, and we will present in the following a completely new, exact and 
convergent technique to extract level density and $\gamma$ strength function 
from primary $\gamma$ spectra. 

\section{Extracting level density and $\gamma$ strength function}

\subsection{Ansatz}

We take the experimental primary $\gamma$ matrix $\Gamma(E_i,E_\gamma)$ as the 
starting point for this discussion. We assume that this matrix is normalized 
for every excitation energy bin $E_i$. This is done by letting the sum of 
$\Gamma$ over all $\gamma$ energies $E_\gamma$ from some minimum $\gamma$ 
energy $E_\gamma^{\mathrm{min}}$ to the maximum $\gamma$ energy $E_i$ at this 
excitation energy bin be unity, i.e.\
\begin{equation}
\label{eq:normalization}
\sum_{E_\gamma=E_\gamma^{\mathrm{min}}}^{E_i}\Gamma(E_i,E_\gamma)=1.
\end{equation}

\begin{figure}[htbp]\centering
\mbox{\psfig{figure=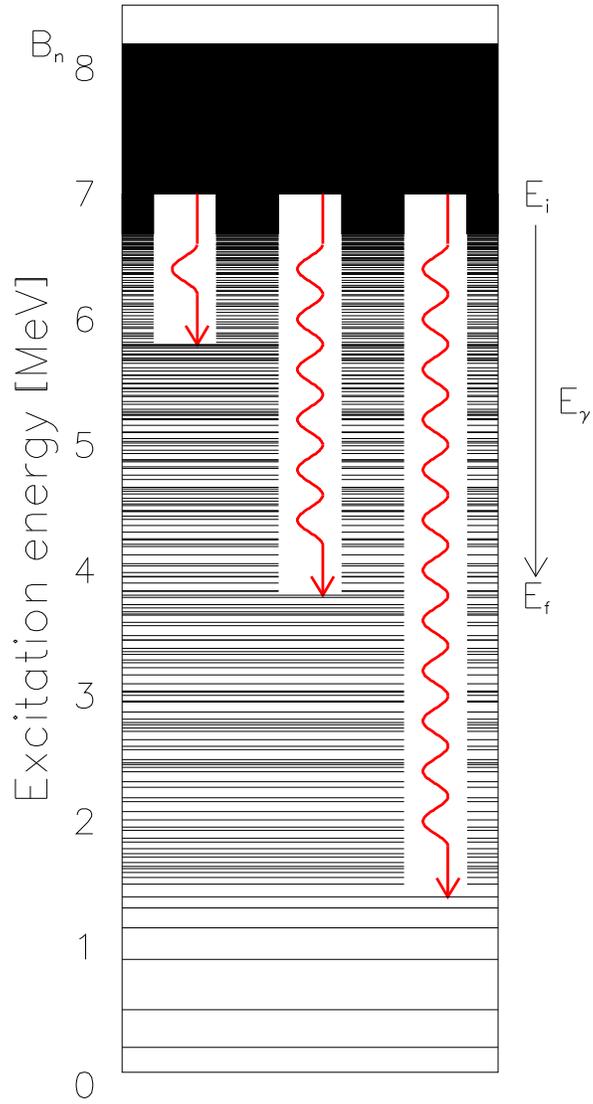,height=16.7cm}}
\caption{The $\gamma$ decay probability from an initial excitation energy $E_i$
in the statistical region is proportional to the level density at the final 
excitation energy $E_f$ and the $\gamma$ strength function at the $\gamma$ 
energy $E_\gamma=E_i-E_f$.}
\label{fig:levels}
\end{figure}

The $\gamma$ decay probability from the excitation energy $E_i$ to $E_f$ by a
$\gamma$ ray with energy $E_\gamma=E_i-E_f$ in the continuum energy region is 
proportional to the level density $\varrho(E_f)$ and a $\gamma$ energy 
dependent factor $F(E_\gamma)$ \cite{Br55,Ax62}. This ansatz is illustrated in
Fig.~\ref{fig:levels}. The experimental normalized primary $\gamma$ matrix 
$\Gamma$ can therefore theoretically be approximated by
\begin{equation}
\label{eq:normansatz}
\Gamma_{\mathrm{th}}(E_i,E_\gamma)=\frac{F(E_\gamma)\,\varrho(E_i-E_\gamma)}
{\sum_{E_\gamma=E_\gamma^{\mathrm{min}}}^{E_i}F(E_\gamma)\,
\varrho(E_i-E_\gamma)},
\end{equation}
which also fulfills Eq.~(\ref{eq:normalization}).

As it is shown in Appendix~\ref{app:proof}, one can construct all solutions of 
Eq.~(\ref{eq:normansatz}) by applying the transformation given by 
Eq.~(\ref{eq:transformation}) to one arbitrary solution, where the generators 
of the transformation $A$, $B$ and $\alpha$ can be chosen freely.

\begin{eqnarray}
\label{eq:transformation}
\tilde{\varrho}(E_i-E_\gamma)&=&\varrho(E_i-E_\gamma)\,
A\exp(\alpha\,(E_i-E_\gamma)),\\
\tilde{F}(E_\gamma)&=&F(E_\gamma)\,B\exp(\alpha\,E_\gamma),\nonumber
\end{eqnarray}

\subsection{Method}

\subsubsection{$0^{\mathrm{th}}$ order estimate}

Since all possible solutions of Eq.~(\ref{eq:normansatz}) can be obtained by 
the transformation given by Eq.~(\ref{eq:transformation}) of one arbitrary 
solution, we choose conveniently $\varrho^{(0)}=1$. With this choice, the 
$0^{\mathrm{th}}$ order estimate of $F$ is given by
\begin{equation}
\Gamma(E_i,E_\gamma)=\frac{F^{(0)}(E_\gamma)}
{\sum_{E_\gamma=E_\gamma^{\mathrm{min}}}^{E_i}F^{(0)}(E_\gamma)}.
\end{equation}
Summing over the excitation energy interval 
$E_i^{\mathrm{min}}\ldots E_i^{\mathrm{max}}$ while obeying $E_i\geq E_\gamma$
yields
\begin{equation}
\sum_{E_i=\max(E_i^{\mathrm{min}},E_\gamma)}^{E_i^{\mathrm{max}}}
\Gamma(E_i,E_\gamma)=F^{(0)}(E_\gamma)\,
\sum_{E_i=\max(E_i^{\mathrm{min}},E_\gamma)}^{E_i^{\mathrm{max}}}\frac{1}
{\sum_{E_\gamma=E_\gamma^{\mathrm{min}}}^{E_i}F^{(0)}(E_\gamma)},
\end{equation}
where the sum on the right hand side can be set to unity, giving
\begin{equation}
F^{(0)}(E_\gamma)=
\sum_{E_i=\max(E_i^{\mathrm{min}},E_\gamma)}^{E_i^{\mathrm{max}}}
\Gamma(E_i,E_\gamma).
\end{equation}

\subsubsection{Higher order estimates}

In order to calculate higher order estimates of the $\varrho$ and $F$ 
functions, we developed a least $\chi^2$ method. The basic idea of this method 
is to minimize
\begin{equation}
\chi^2=\frac{1}{N_{\mathrm{free}}}\,
\sum_{E_i=\max(E_i^{\mathrm{min}},E_\gamma)}^{E_i^{\mathrm{max}}}
\sum_{E_\gamma=E_\gamma^{\mathrm{min}}}^{E_i}
\left(\frac{\Gamma_{\mathrm{th}}(E_i,E_\gamma)-\Gamma(E_i,E_\gamma)}
{\Delta\Gamma(E_i,E_\gamma)}\right)^2,
\end{equation}
where $N_{\mathrm{free}}$ is the number of degrees of freedom, and 
$\Delta\Gamma(E_i,E_\gamma)$ is the uncertainty in the primary $\gamma$ matrix.
Since we assume every point of the $\varrho$ and $F$ functions as independent 
variables, we calculate $N_{\mathrm{free}}$ as
\begin{equation}
N_{\mathrm{free}}={\mathrm{ch}}(\Gamma)-{\mathrm{ch}}(\varrho)-
{\mathrm{ch}}(F),
\end{equation}
where ch indicates the number of data points in the respective spectra. 

We minimize the reduced $\chi^2$ by letting all derivatives 
\begin{equation}
\label{eq:derivatives}
\frac{\partial}{\partial\,F(E_\gamma)}\chi^2=0
\hspace*{1cm}{\mathrm{and}}\hspace*{1cm}
\frac{\partial}{\partial\,\varrho(E_i-E_\gamma)}\chi^2=0
\end{equation}
for every argument $E_\gamma$ and $E_i-E_\gamma$ respectively. A rather tedious
but straight forward calculation yields equivalence between 
Eqs.~(\ref{eq:derivatives}) and
\begin{eqnarray}
\label{eq:iterationF}
F(E_\gamma)&=&\frac{
\sum_{E_i=\max(E_i^{\mathrm{min}},E_\gamma)}^{E_i^{\mathrm{max}}}
\varrho(E_i-E_\gamma)\,\varphi(E_i,E_\gamma)}
{\sum_{E_i=\max(E_i^{\mathrm{min}},E_\gamma)}^{E_i^{\mathrm{max}}}
\varrho^2(E_i-E_\gamma)\,\psi(E_i,E_\gamma)}\\
\label{eq:iterationrho}
\varrho(E_f)&=&\frac{\sum_{E_i=\max(E_i^{\mathrm{min}},E_f+
E_\gamma^{\mathrm{min}})}^{E_i^{\mathrm{max}}}
F(E_i-E_f)\,\varphi(E_i,E_i-E_f)}
{\sum_{E_i=\max(E_i^{\mathrm{min}},E_f+
E_\gamma^{\mathrm{min}})}^{E_i^{\mathrm{max}}}
F^2(E_i-E_f)\,\psi(E_i,E_i-E_f)},
\end{eqnarray}
where
\begin{eqnarray}
\varphi(E_i,E_\gamma)&=&\frac{a(E_i)}{s^3(E_i)}-\frac{b(E_i)}{s^2(E_i)}
+\frac{\Gamma(E_i,E_\gamma)}{s(E_i)\,
\left(\Delta\Gamma(E_i,E_\gamma)\right)^2}\\
\psi(E_i,E_\gamma)&=&\frac{1}{\left(s(E_i)\,\Delta\Gamma(E_i,E_\gamma)
\right)^2},
\end{eqnarray}
and
\begin{eqnarray}
a(E_i)&=&\sum_{E_\gamma=E_\gamma^{\mathrm{min}}}^{E_i}
\left(\frac{F(E_\gamma)\,\varrho(E_i-E_\gamma)}
{\Delta\Gamma(E_i,E_\gamma)}\right)^2\\
b(E_i)&=&\sum_{E_\gamma=E_\gamma^{\mathrm{min}}}^{E_i}
\frac{F(E_\gamma)\,\varrho(E_i-E_\gamma)\,\Gamma(E_i,E_\gamma)}
{\left(\Delta\Gamma(E_i,E_\gamma)\right)^2}\\
s(E_i)&=&\sum_{E_\gamma=E_\gamma^{\mathrm{min}}}^{E_i}
F(E_\gamma)\,\varrho(E_i-E_\gamma).
\end{eqnarray}

\begin{figure}[htbp]\centering
\mbox{\psfig{figure=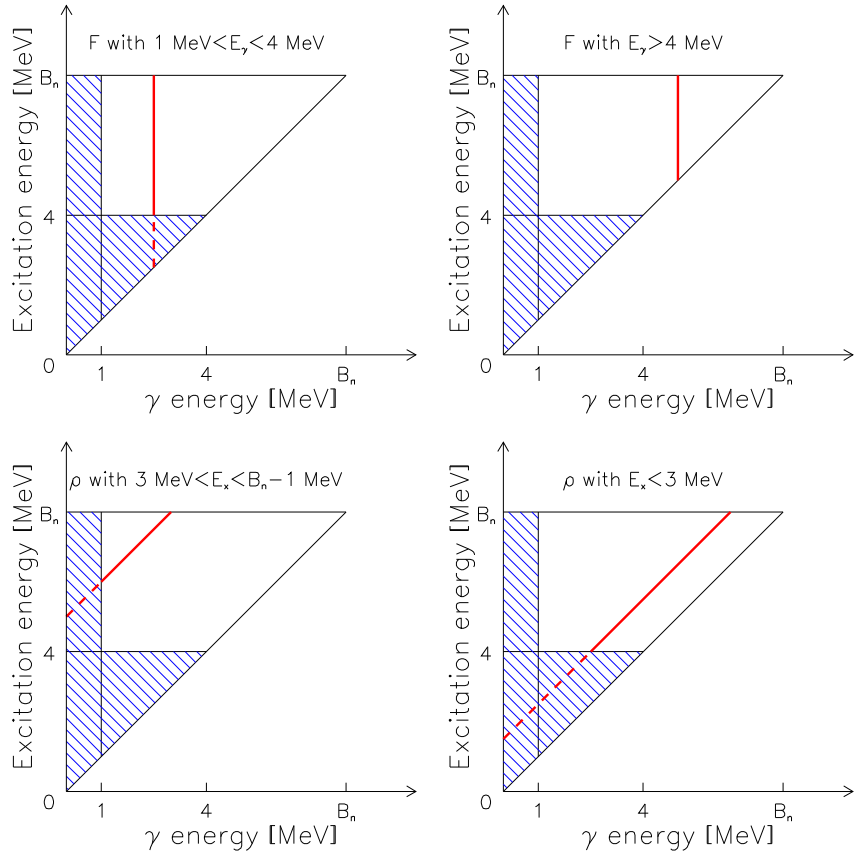,height=12.1cm}}
\caption{Performing the sum of Eqs.~(\protect{\ref{eq:iterationF}}) and 
(\protect{\ref{eq:iterationrho}}) along the thick lines. The shaded areas are
usually excluded when extracting real data due to methodical problems (see 
text).}
\label{fig:method}
\end{figure}

Within one iteration, we first calculate the functions $a(E_i)$, $b(E_i)$ and 
$s(E_i)$, using the previous order estimates for $\varrho$ and $F$. Using these
three functions, we can calculate the matrices $\varphi(E_i,E_\gamma)$ and 
$\psi(E_i,E_\gamma)$. Further on, we calculate the actual order estimates of 
$\varrho$ and $F$ by means of Eqs.~(\ref{eq:iterationF}) and 
(\ref{eq:iterationrho}). Figure~\ref{fig:method} shows where the sums in
Eqs.~(\ref{eq:iterationF}) and (\ref{eq:iterationrho}) are performed.

\subsubsection{Convergence properties}

The method usually converges very well. However, in some cases the $\chi^2$
minimum is very shallow, and the chance exists, that the iteration procedure
might fail. In order to enhance convergence of the method, we have restricted 
the maximum change of every data point in $\varrho$ and $F$ within one 
iteration to a certain percentage $P$. This means that the data point obtained 
in the actual iteration (new) is checked if it lies within the interval 
\begin{equation}
\label{eq:interval}
\frac{\mathrm{old}}{(1+P/100)}\leq{\mathrm{new}}\leq (1+P/100)\cdot
{\mathrm{old}},
\end{equation}
determined by the data point from the previous iteration (old). In the case 
that the new data point lies outside this interval, it will be set to the value
of the closest boundary. 

Applying this method to some of our data, we have observed, that the smaller
$P$ is chosen, the smaller $\chi^2$ gets in the end, when the procedure has
reached its limit. The reason for this is, that more and more data points
in $\varrho$ and $F$ will converge, while fewer and fewer points (typically
at high energies $E_\gamma$ and $E_f$ where few counts are available) are 
oscillating between the two boundaries given by Eq.~(\ref{eq:interval}). 
Occasionally, we can choose $P$ so small that all data points will converge and
no oscillating behavior can be seen. However, in some cases oscillating data 
points can not be avoided by any choice of $P$ which might indicate that the 
$\chi^2$ minimum is too shallow, or does not even exist, for some data points 
in $\varrho$ and $F$.

A small $P$ would lead to an accurate result but make a large number of 
iterations necessary, and a large $P$ would shorten the execution time of the
procedure but affect the accurateness of the solution. We combine the 
advantages and avoid the disadvantages of the two concepts by letting $P$ 
become smaller as a function of the number of iterations. In our actual 
computer code \cite{SB99b}, we have implemented a stepwise decrease of $P$ as 
shown in Table~\ref{tab:decrease}. The choices of $P$ as a function of the 
number of iterations is quite arbitrary, but we have achieved very good 
convergence for those spectra, where convergence properties without 
restrictions are rather fair. 

In conclusion we have to stress, that the convergence properties of the method
in many cases do not require any restrictions of the maximum variation of data
points within one iteration. In those cases however, where restrictions are 
mandatory to achieve or enhance convergence, they will only affect a small 
percentage of the data points at high energies, where data in the primary 
$\gamma$ matrix are sparse and mainly erratically scattered. In those cases, 
where the restrictions of Table~\ref{tab:decrease} would prove not to be 
satisfactory for convergence, the number of iterations or the value of $P$ 
can be changed, since the validity of the method does not rely on these values.

\subsubsection{Error calculation}

A huge effort has been made in order to estimate errors of the data points in
$\varrho$ and $F$. Since the experimental primary $\gamma$ matrix has been 
obtained from raw data by applying an unfolding procedure \cite{GT96} and a 
subtraction technique \cite{GR87}, error propagation through these methods is 
very tedious and has never been performed. In order to perform an error 
estimation of $\varrho$ and $F$, we first have to estimate the error of the
primary $\gamma$ matrix data. A rough estimation yields
\begin{equation}
\label{eq:1error}
\Delta\Gamma=2\,\sqrt{(M_1+M_2)\,\Gamma},
\end{equation}
where $M_1$ denotes the number of first and higher generation $\gamma$ rays, 
and $M_2$ the number of second and higher generation $\gamma$ rays at one 
excitation energy bin $E_i$. We estimate those quantities roughly by
\begin{equation}
M_1=\max(1,M(E_i))
\hspace*{1cm}{\mathrm{and}}\hspace*{1cm}
M_2=\max(0,M(E_i)-1),
\end{equation}
where the multiplicity $M(E_i)$ is given by a fit to the experimental data in 
Ref.~\cite{TB95}
\begin{equation}
M(E_i)=0.42+4.67\,10^{-4}\,E_i-1.29\,10^{-8}\,E_i^2,
\end{equation}
and $E_i$ is given in keV. The motivation of Eq.~(\ref{eq:1error}) is that 
during the extraction method of primary $\gamma$ spectra of Ref.~\cite{GR87} 
the second and higher generation $\gamma$ ray spectrum, which has of the order 
$M_2\,\Gamma$ counts, is subtracted from the total unfolded $\gamma$ ray 
spectrum, which has of the order $M_1\,\Gamma$ counts. The errors of these 
spectra are roughly the square root of the number of counts. If we assume that 
these errors are independent from each other, the primary $\gamma$ spectra has 
an error of roughly $\sqrt{(M_1+M_2)\,\Gamma}$. The factor 2 in 
Eq.~(\ref{eq:1error}) is due to the unfolding procedure and is quite uncertain.
We assume this factor to be roughly equal the ratio of the solid angle covered
by the CACTUS detector array of some 15\% \cite{GA90} to its photopeak 
efficiency of some 7\% at 1.3~MeV \cite{SB98}. We have however to apply a 
couple of minor corrections to Eq.~(\ref{eq:1error}). 

Firstly, the first generation method \cite{GR87} exhibits some methodical
problems at low excitation energies. The basic assumption behind this method is
that the $\gamma$ decay properties of an excited state is unaffected by its
formation mechanism e.g.\ direct population by a nuclear reaction, or 
population by a nuclear reaction followed by one or several $\gamma$ rays. This
assumption is not completely valid at low excitation energies, where 
thermalization time might compete with the half life of the excited state and
the reactions used exhibit a more direct than compound character. This and some
experimental problems like ADC threshold walk and bad timing properties of 
low energetic $\gamma$ rays, all described in Ref.~\cite{SB98}, oblige us to 
exclude $\gamma$ rays below 1~MeV from the primary $\gamma$ spectra. For 
low energetic $\gamma$ rays above 1~MeV, we increase the error bars by the 
following rule. For each excitation energy bin $E_i$, we identify the channel 
with the maximum number of counts ch$^{\mathrm{max}}$ (this occurs typically 
between 2 and 3~MeV of $\gamma$ energy). This is also the channel with the 
highest error err$^{\mathrm{max}}$, following Eq.~(\ref{eq:1error}). We then 
replace the errors of the channels ch below ch$^{\mathrm{max}}$ by 
\begin{equation}
{\mathrm{err}}={\mathrm{err}}^{\mathrm{max}}\,\left(1+1.0\cdot
\frac{{\mathrm{ch}}^{\mathrm{max}}-{\mathrm{ch}}}{{\mathrm{ch}}^{\mathrm{max}}}
\right).
\end{equation}
This formula cannot be motivated by some simple handwaving arguments. We feel
however, after inspecting several primary $\gamma$ matrices, that we estimate
the systematic error of these spectra quite accurate.     

Secondly, the unfolding procedure \cite{GT96} exhibits some methodical problems
at high $\gamma$ energies. Since the ratio of the photopeak efficiency to the
solid angle covered by the CACTUS detector array drops for higher $\gamma$
energies, the counts at these energies are multiplied with significant factors 
in the unfolding procedure. Some channels might nevertheless turn out to 
contain almost zero counts, giving differences in counts between two 
neighboring channels by two orders of magnitude. Since the errors are estimated
as proportional to the square root of the number of counts, the estimated 
errors of these channels do not reflect their statistical significance. In 
order to obtain comparable errors to neighboring channels we check the errors 
within one excitation energy bin from the $\gamma$ energy of $\sim$4~MeV and 
upwards. If the error drops by more than a factor 2, when going from one 
channel to the next higher one, we set the error of the higher channel equal to
50\% of the error of the previous one. Also this rule cannot be motivated by a 
simple argumentation. It affects, however usually only a very small percentage 
of channels, and an inspection of several primary $\gamma$ spectra gives us 
confidence in our error estimation.

It is now very tedious to perform error propagation calculation through the
extraction procedure. We therefore decided to apply a simulation technique to
obtain reliable errors of the $\varrho$ and $F$ functions. For this reason, we 
add statistical fluctuations to the primary $\gamma$ matrix. For every channel
in the primary $\gamma$ matrix, we choose a random number $r$ between zero and 
one. We then calculate $x$ according to 
\begin{equation}
r=\frac{1}{\sqrt{2\pi}\,\sigma}\int_{-\infty}^x\exp(-\frac{(\xi-a)^2}
{2\,\sigma^2})\,{\mathrm{d}}\xi,
\end{equation}   
where $a$ is the number of counts and $\sigma$ the error of this channel. By 
replacing the number of counts $a$ with $x$, we add a statistical fluctuation 
to this channel. This is done for all channels of the primary $\gamma$ matrix, 
and new $\varrho^{(s)}$ and $F^{(s)}$ functions are extracted, containing 
statistical fluctuations. This procedure is repeated 100 times, which gives
reasonable statistics. The errors in $\varrho$ and $F$ are then calculated by
\begin{eqnarray}
\Delta\varrho(E_f)&=&\frac{1}{\sqrt{100}}\,\sqrt{\sum_{i=1}^{100}
[\varrho_i^{(s)}(E_f)-\varrho(E_f)]^2}\\
\Delta F(E_\gamma)&=&\frac{1}{\sqrt{100}}\,\sqrt{\sum_{i=1}^{100}
[F_i^{(s)}(E_\gamma)-F(E_\gamma)]^2}.
\end{eqnarray}

\subsubsection{Normalizing the level density to other experimental data}
\label{sec:ambiguity}

As pointed out above, all solutions of Eq.~(\ref{eq:normansatz}) can be 
generated from one arbitrary solution by the transformation given by 
Eq.~(\ref{eq:transformation}). It is of course discouraging that an infinite 
number of equally good solutions exists, however by comparing to known data, 
we will be able to pick out the most physical one. 

At low excitation energies up to typically 2~MeV for even even nuclei, we can 
compare the extracted level density to the number of known levels per 
excitation energy bin (for a comprehensive compilation of all known levels in  
nuclei see e.g.\ Ref.~\cite{FS96}). At the neutron binding energy, we can 
deduce the level density for many nuclei from available neutron resonance 
spacing data. The starting point is Eqs.~(4) and (5) of Ref.~\cite{GC65}
\begin{eqnarray}
\label{eq:leveldensityspin}
\varrho(U,J)&=&\frac{\sqrt{\pi}}{12}\ \frac{\exp 2\sqrt{aU}}{a^{1/4}U^{5/4}}\ 
\frac{(2J+1)\exp(-(J+1/2)^2/2\sigma^2)}{2\sqrt{2\pi}\sigma^3}\\
\label{eq:leveldensity}
\varrho(U)&=&\frac{\sqrt{\pi}}{12}\ \frac{\exp 2\sqrt{aU}}{a^{1/4}U^{5/4}}\ 
\frac{1}{\sqrt{2\pi}\sigma},
\end{eqnarray} 
where $\varrho(U,J)$ is the level density for both parities and for a given 
spin $J$, and $\varrho(U)$ is the level density for all spins and parities; 
$\sigma$ is the spin dependence parameter and $a$ the level density parameter.
Assuming that $I$ is the spin of the target nucleus in a neutron resonance 
experiment, the neutron resonance spacing $D$ can be written as
\begin{equation}
\label{eq:spacing}
\frac{1}{D}=\frac{1}{2}(\varrho(U_n,J=I+1/2)+\varrho(U_n,J=I-1/2)),
\end{equation}
since all levels are accessible in neutron resonance experiments, and we 
assume, that both parities contribute equally to the level density at the 
neutron binding energy represented by $U_n$. Combining 
Eqs.~(\ref{eq:leveldensityspin}), (\ref{eq:leveldensity}) and 
(\ref{eq:spacing}), one can calculate the total level density at the neutron 
binding energy
\begin{equation}
\varrho(U_n)=\frac{2\sigma^2}{D}\ \frac{1}{(I+1)\exp(-(I+1)^2/2\sigma^2)
+I\exp(-I^2/2\sigma^2)},
\end{equation}
where $\sigma^2$ is calculated by combining Eqs.~(9) and (11) of 
Ref.~\cite{GC65} i.e.\ 
\begin{equation}
\label{eq:sigma}
\sigma^2=0.0888\sqrt{aU_n}A^{2/3},
\end{equation}
and $A$ is the mass number of the nucleus. It is assumed that $\sigma^2$ has 
an error of $\sim$10\% due to shell effects \cite{GC65}. One should also point 
out, that $U_n$ is given by $U_n=B_n-P$, where $B_n$ is the neutron binding 
energy and $P$ the pairing energy which can be found in Table III of 
Ref.~\cite{GC65} for many nuclei. Unfortunately, we cannot compare the 
calculated level density at the neutron binding energy directly with our 
extracted level density, since due to the omission of $\gamma$ rays below 
1~MeV, the $\varrho$ function can only be extracted up to 1~MeV below the 
neutron binding energy. We will however extrapolate the extracted $\varrho$ 
function with a Fermi gas level density, obtained by combining 
Eqs.~(\ref{eq:leveldensity}) and (\ref{eq:sigma})
\begin{equation}
\label{eq:extrapolation}
\varrho(U)=\frac{1}{12\sqrt{0.1776}A^{1/3}}\ \frac{\exp 2\sqrt{aU}}
{a^{1/2}U^{3/2}}.
\end{equation}
This is done by adjusting the parameters $A$ and $\alpha$ of the transformation
given by Eq.~(\ref{eq:transformation}) such, that the data fit the level 
density formula of Eq.~(\ref{eq:extrapolation}) in an excitation energy 
interval between 3 and 1~MeV below $B_n$, where in most cases all parameters of
Eq.~(\ref{eq:extrapolation}) can be taken from Tables II and III of 
Ref.~\cite{GC65}. This semi experimental level density spanning from 0~MeV up 
to $B_n$ is then again transformed according to Eq.~(\ref{eq:transformation}) 
such, that it fits the number of known levels up to $\sim$2~MeV and $\sim$1~MeV
for even even and odd even nuclei respectively and simultaneously the level 
density deduced from neutron resonance spacing at $B_n$. We have to point out 
however, that after the fit to known data, the extrapolation does not have the 
functional form of Eq.~(\ref{eq:extrapolation}) anymore, due to the 
transformation given by Eq.~(\ref{eq:transformation}) applied to the semi 
experimental level density. Therefore, if necessary, a new extrapolation of the
experimental data must be performed. 

We have successfully implemented the new extraction method in a Fortran 77 
computer code called \sc rhosigchi \rm\cite{SB99b}. The computer code was 
compiled under a Solaris 2.5.1 operating system running on a Dual UltraSPARC 
station with 200~MHz CPU. The execution time of one extraction is in the order
of 10-20~s. The computer code has $\sim$1200 programming lines, excluding 
special library in and output routines.  

\section{Applications to spectra}

\subsection{Testing the method on theoretical spectra}

\begin{figure}[htbp]\centering
\mbox{\psfig{figure=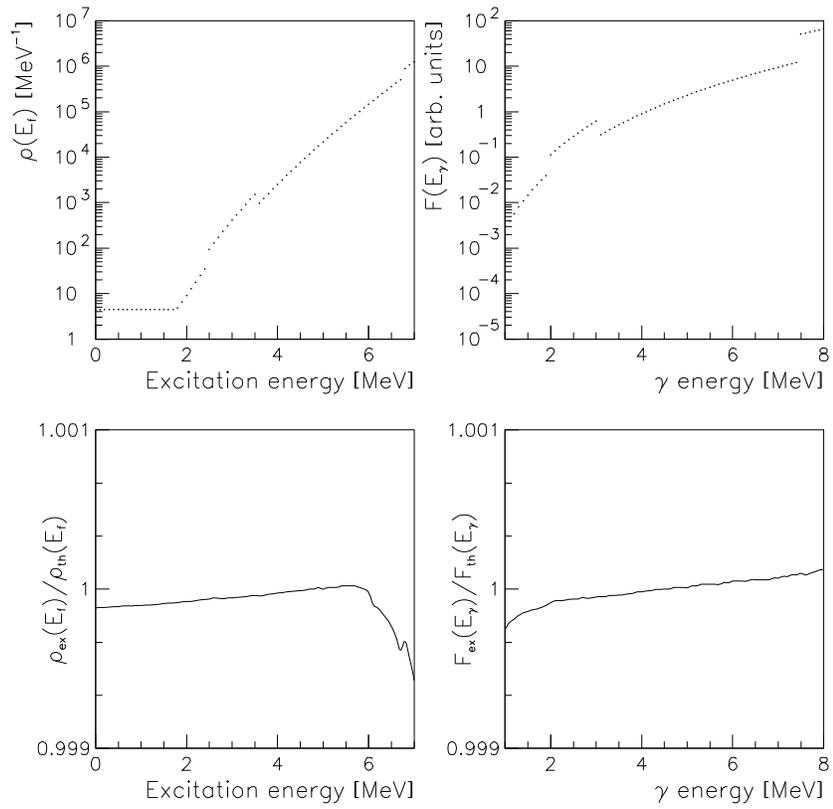,height=12.1cm}}
\caption{Theoretical level density $\varrho$ and $\gamma$ energy dependent
factor $F$ used to calculate a primary $\gamma$ matrix (upper half). Ratio of 
extracted to theoretical $\varrho$ and $F$ functions (lower half).}
\label{fig:test}
\end{figure}

The method has been tested on a theoretically calculated primary $\gamma$ 
matrix. The theoretical primary $\gamma$ matrix was obtained by simply 
multiplying a level density $\varrho$ to a $\gamma$ energy dependent factor $F$
according to Eq.~(\ref{eq:normansatz}). The level density was given by a 
backshifted Fermi gas formula
\begin{equation}
\varrho(U)=C\,U^{-3/2}\,\exp(2\sqrt{aU})
\end{equation}
with $U=E_f-P$. Below the minimum at $U=9/4a$ a constant level density was 
used. The $\gamma$ energy dependent factor was chosen as
\begin{equation}
F(E_\gamma)=C\,E_\gamma^{4.2}.
\end{equation}
In addition, a ``fine structure'' was imposed on both functions, by scaling
several $\sim$1~MeV broad intervals with factors around 1.5--5. Both model
functions are shown in the upper half of Fig.~\ref{fig:test}. We extracted the 
$\varrho$ and $F$ functions from the theoretical primary $\gamma$ matrix using
the excitation energy interval of 4 to 8~MeV and excluding all $\gamma$ rays 
below 1~MeV. In the lower panel of Fig.~\ref{fig:test} we show the ratio of the
extracted functions to the theoretical functions. After adjusting the extracted
quantities with the transformation given by Eq.~(\ref{eq:transformation}), we 
can state that the deviation from the input functions is smaller than one per 
thousand in the covered energy range of both functions. Tests of the old 
extraction method showed deviations of the order of 10\% to 100\% \cite{ST97}. 
We therefore consider the new extraction method to be much more reliable. 

\subsection{Testing the method on $^{172}$Yb spectra}

We have tested the method on several experimental primary $\gamma$ spectra. We
will in the following discuss a typical example; the 
$^{173}$Yb($^3$He,$\alpha$)$^{172}$Yb reaction. The experiment was carried out
at the Oslo Cyclotron Laboratory (OCL) at the University of Oslo, using a MC35
cyclotron with a beam energy of 45~MeV and a beam intensity of typically 
1~nA. The experiment was running for two weeks. The target was consisting of a 
self supporting, isotopically enriched (92\% $^{173}$Yb) metal foil of 
2.0~mg/cm$^2$ thickness. Particle identification and energy measurements were 
performed by a ring of 8 Si(Li) particle telescopes at 45$^\circ$ with respect 
to the beam axis. The $\gamma$ rays were detected by an array of 28 
$5''\times5''$ NaI(Tl) detectors (CACTUS). More experimental details can be 
found in \cite{GA90}. The raw data are unfolded, using measured response 
functions of the CACTUS detector array \cite{GT96}. After unfolding, a 
subtraction method is applied to the particle $\gamma$ matrix in order to 
extract the first generation $\gamma$ matrix \cite{GR87}. This primary $\gamma$
matrix is taken as the starting point for the extraction method presented here.

\begin{figure}[htbp]\centering
\mbox{\psfig{figure=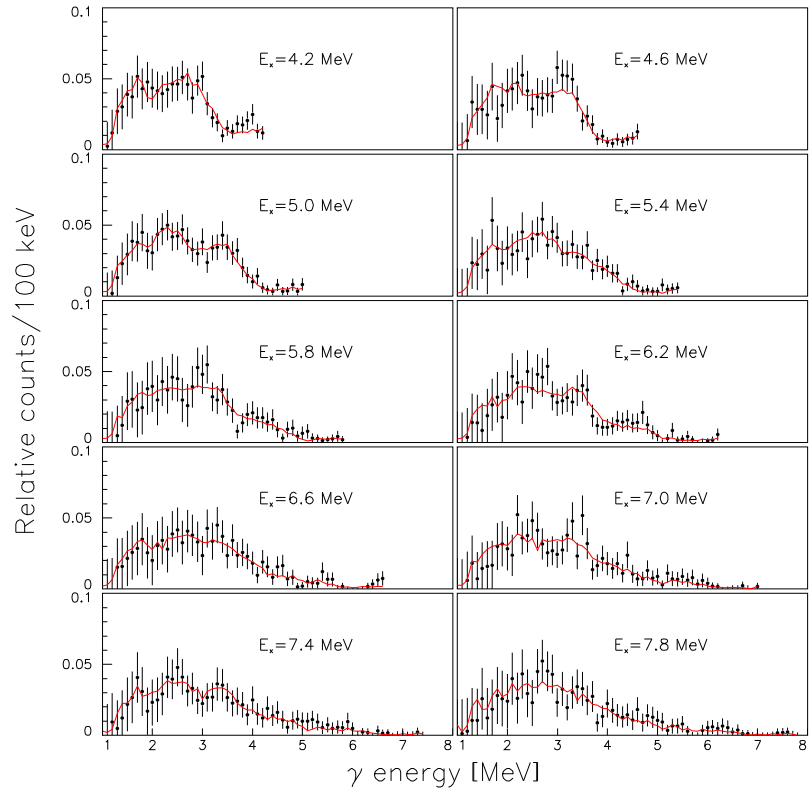,height=12.1cm}}
\caption{Normalized experimental primary $\gamma$ spectra with estimated errors
(data points) and calculated primary $\gamma$ spectra from the extracted level 
density $\varrho$ and $\gamma$ energy dependent factor $F$ according to 
Eq.~(\protect{\ref{eq:normansatz}}) (lines).}
\label{fig:fgboth}
\end{figure}

In Fig.~\ref{fig:fgboth}, we show the normalized, experimental primary $\gamma$
spectra at ten different excitation energy bins (data points). The errors of 
the data points are estimated as explained above. The $\varrho$ and $F$ 
functions were extracted from the excitation energy interval 4-8~MeV, excluding
all $\gamma$ energies smaller than 1~MeV. The lines are the calculated 
primary $\gamma$ spectra, obtained by multiplying the extracted level density 
$\varrho$ and the $\gamma$ energy dependent factor $F$ according to 
Eq.~(\ref{eq:normansatz}). One can see, that the lines follow the data points 
very well. It can also be seen that the errors of the data points are estimated
reasonably giving a reduced $\chi^2$ of $\sim$0.4. Figure~\ref{fig:fgboth} is a
beautiful example for the claim, that primary $\gamma$ spectra can be 
factorized according to the Axel Brink hypothesis \cite{Br55,Ax62}.

\begin{figure}[htbp]\centering
\mbox{\psfig{figure=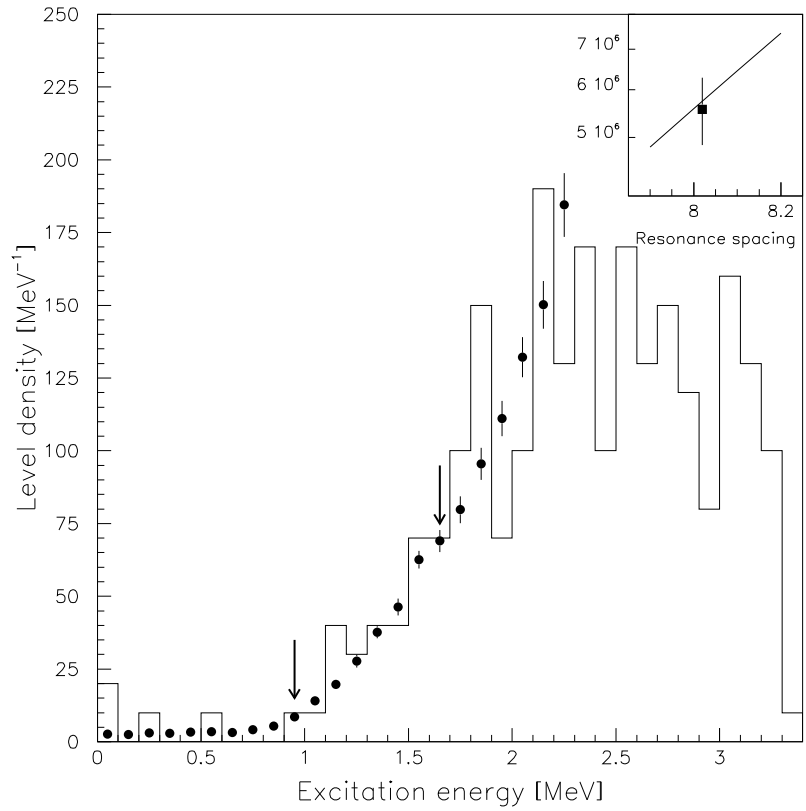,height=12.1cm}}
\caption{Picking out the most physical solution of 
Eq.~(\protect{\ref{eq:normansatz}}) by comparing the extracted level density 
$\varrho$ (data points) to the number of known levels per excitation energy bin
(histogram) between the arrows, and the extrapolation of $\varrho$ up to $B_n$
(line in insert, see text) to the neutron resonance spacing data (data point in
insert, see text).}
\label{fig:counting}
\end{figure}

Figure~\ref{fig:counting} shows how the parameters $\alpha$ and $A$ of the
transformation given by Eq.~(\ref{eq:transformation}) can be determined in the 
case of the $^{173}$Yb($^3$He,$\alpha$)$^{172}$Yb reaction. The extracted 
$\varrho$ function (data points) is compared to the number of known levels 
\cite{FS96} per excitation energy bin (histogram) and to the level density at 
the neutron binding energy, calculated from neutron resonance spacing data 
\cite{LC73} (data point in insert). The line in the insert is the extrapolation
of the $\varrho$ function up to $B_n$ according to Section \ref{sec:ambiguity}.

\begin{figure}[htbp]\centering
\mbox{\psfig{figure=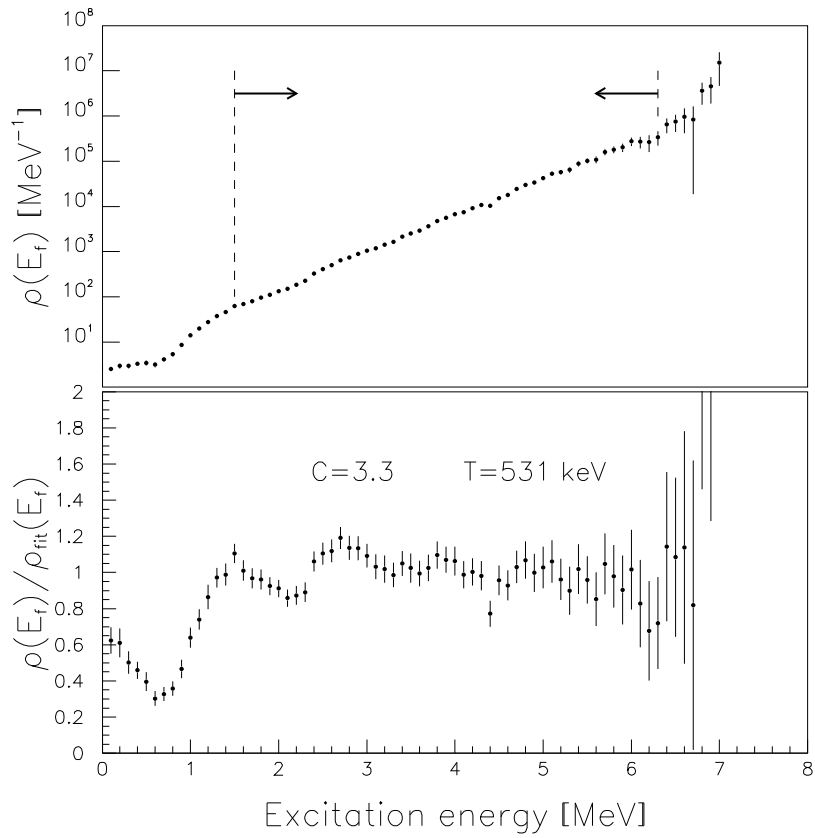,height=12.1cm}}
\caption{Extracted normalized level density in $^{172}$Yb (upper panel), and
the same data divided by an exponential fit (lower panel, see text).}
\label{fig:rho}
\end{figure}

In the following, the extracted $\varrho$ and $F$ functions are discussed. Both
functions were already published before, using the old extraction method and 
some fine structure discussed below, could already be seen in the previous 
publication \cite{TB96}. Figure~\ref{fig:rho} shows the level density $\varrho$
and the relative level density, which is the level density, divided by an 
exponential fit to the data between the arrows. The parameters of the fit 
function 
\begin{equation}
\varrho_{\mathrm{fit}}=C\exp(E/T)
\end{equation}
are shown in the lower panel of the figure. In the relative level density one 
can see a small bump emerging at $\sim$2.7~MeV probably due to the quenching of
pairing correlations \cite{MB99,SB99}. One can also see very nicely the onset 
of strong pairing correlations at 1.0--1.5~MeV of excitation energy. 

\begin{figure}[htbp]\centering
\mbox{\psfig{figure=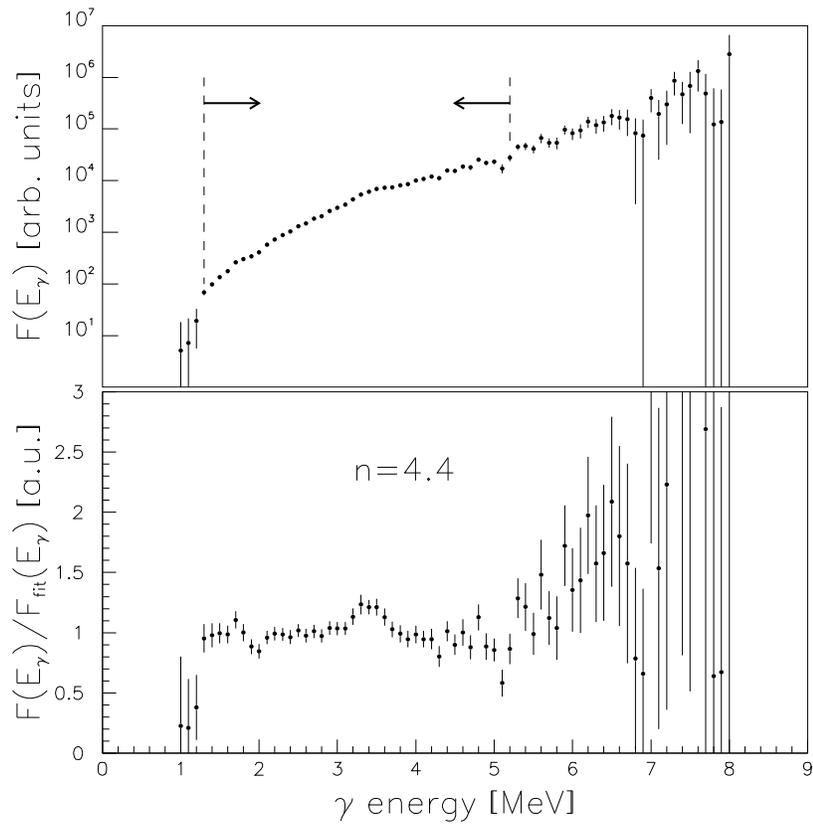,height=12.1cm}}
\caption{Extracted $\gamma$ energy dependent factor $F$ in $^{172}$Yb (upper 
panel), and the same data divided by a fit function (lower panel, see text).}
\label{fig:sig}
\end{figure}

In Fig.~\ref{fig:sig} the $\gamma$ energy dependent factor is shown (upper 
panel). On the lower panel, the same data are given, divided by a fit function 
of the form
\begin{equation}
F_{\mathrm{fit}}=C\,E_\gamma^n.
\end{equation}
This function can be used as a parameterization of 
\begin{equation}
F(E_\gamma)=E_\gamma^{2\lambda+1}\,\sigma(E_\gamma),
\end{equation}
where $\sigma(E_\gamma)$ is the $\gamma$ strength function and $\lambda$ is the
multipolarity of the $\gamma$ transition. The fit to the data was performed 
between the arrows, the fit parameter $n$ is given in the lower panel. Since 
other experimental data is very sparse, we did not scale $F$ in order to obtain
absolute units. However, the extracted fit parameter $n$ is in good agreement 
with expectations from the tail of a GDR strength function at low $\gamma$ 
energies \cite{HB95}. In the lower panel a merely significant bump at 
$\sim$3.4~MeV is visible, which we interpret as the Pigmy resonance.

\section{Conclusions}

In this work we have presented for the first time a reliable and convergent 
method to extract consistently and simultaneously the level density $\varrho$ 
and the $\gamma$ energy dependent function $F$ from primary $\gamma$ spectra. 
The new method, based on a least square fit, has been carefully tested on 
simulated $\gamma$ spectra as well as on experimental data. In order to 
normalize the data, we count known discrete levels in the vicinity of the 
ground state and use the level spacing known from neutron resonances at the 
neutron binding energy.

Compared to the previous projection method \cite{TB96}, the least square fit 
method gives the following advantages: The iteration converges mathematically.
The reproduction of the input level densities and $\gamma$ strength functions 
in simulations is much better (almost exact). No tuning of the initial trial 
function is necessary to obtain a reasonable scaled level density, but the 
newly derived transformation properties of the solution enable the user to 
normalize the extracted quantities with known data. The reduced $\chi^2$ is
estimated reasonably. The errors of the extracted quantities are estimated by 
statistical simulations.

We have used the new method to reanalyze previously published data and for the
analysis of more recent data. Especially the ability to extract absolute
values of the level density $\varrho$ enables us to perform several new
applications \cite{SB99,GH99,GB99}.

\section{Acknowledgments}

The authors wish to thank A.~Bjerve for interesting discussions. Financial 
support from the Norwegian Research Council (NFR) is gratefully acknowledged. 

\appendix

\section{Proof of Eq.~(\protect{\ref{eq:transformation}})}
\label{app:proof}

The functional form of Eq.~(\ref{eq:normansatz}) opens for a manifold of 
solutions. If one solution of Eq.~(\ref{eq:normansatz}) is found, one can 
generally construct all possible solutions by the following transformation
\begin{eqnarray}
\label{eq:generaltransformation}
\tilde{\varrho}(E_i-E_\gamma)&=&\varrho(E_i-E_\gamma)\,g(E_i-E_\gamma),\\
\tilde{F}(E_\gamma)&=&F(E_\gamma)\,f(E_\gamma).\nonumber
\end{eqnarray}
The two functions $g$ and $f$ have to fulfill certain conditions, since the set
of functions $\tilde{\varrho}$ and $\tilde{F}$ are supposed to form a solution 
of Eq.~(\ref{eq:normansatz}) i.e.
\begin{equation}
\Gamma_{\mathrm{th}}(E_i,E_\gamma)=
\frac{F(E_\gamma)\,\varrho(E_i-E_\gamma)}
{\sum_{E'_\gamma=E_\gamma^{\mathrm{min}}}^{E_i}
F(E'_\gamma)\,\varrho(E_i-E'_\gamma)}=
\frac{\tilde{F}(E_\gamma)\,\tilde{\varrho}(E_i-E_\gamma)}
{\sum_{E'_\gamma=E_\gamma^{\mathrm{min}}}^{E_i}
\tilde{F}(E'_\gamma)\,\tilde{\varrho}(E_i-E'_\gamma)}.
\end{equation}
Inserting Eq.~(\ref{eq:generaltransformation}) one can easily deduce 
\begin{eqnarray}
\lefteqn{f(E_\gamma)\,g(E_i-E_\gamma)\,
\sum_{E'_\gamma=E_\gamma^{\mathrm{min}}}^{E_i}\,
F(E'_\gamma)\,\varrho(E_i-E'_\gamma)=}\\
&&\sum_{E'_\gamma=E_\gamma^{\mathrm{min}}}^{E_i}\,
f(E'_\gamma)\,g(E_i-E'_\gamma)\,F(E'_\gamma)\,\varrho(E_i-E'_\gamma).\nonumber
\end{eqnarray}
Since the right side is independent of $E_\gamma$, also the left side must be
independent of $E_\gamma$, thus the product of $f$ and $g$ must be a function
of $E_i$ only yielding
\begin{equation}
\label{eq:condition}
f(E_\gamma)\,g(E_i-E_\gamma)=h(E_i).
\end{equation}
This condition must of course hold for the case $E_i=E_\gamma$. Using the 
short hand notation $g(0)=A$, one obtains
\begin{equation}
A\,f(E_\gamma)=h(E_\gamma).
\end{equation}
Inserting this result in Eq.~(\ref{eq:condition}), one gets
\begin{equation}
\label{eq:2condition}
f(E_\gamma)\,g(E_i-E_\gamma)=A\,f(E_i).
\end{equation}
Analogously, the condition must hold for the case $E_\gamma=0$, and with 
$f(0)=B$, one obtains
\begin{equation}
\label{eq:determine}
B\,g(E_i)=A\,f(E_i).
\end{equation}
Inserting this result in Eq.~(\ref{eq:2condition}), one finally gets
\begin{equation}
\label{eq:exponential}
g(E_\gamma)\,g(E_i-E_\gamma)=A\,g(E_i).
\end{equation}
We will now show, that the only solution of Eq.~(\ref{eq:exponential}) is an
exponential function. This proof will involve the limit of 
Eq.~(\ref{eq:exponential}) for small $E_\gamma$. However, since $g$ is a 
function of only one variable and the variable $E_i$ is unrestricted in the 
proof, it will be valid for all arguments of $g$. 

By expanding $g$ in Taylor series up to the first order in $E_\gamma$, one 
obtains
\begin{equation}
[A+g'(0)\,E_\gamma]\,[g(E_i)-g'(E_i)\,E_\gamma]=A\,g(E_i).
\end{equation}
Neglecting second order terms in $E_\gamma$ and dividing by $E_\gamma$ one gets
\begin{equation}
g'(0)\,g(E_i)=A\,g'(E_i).
\end{equation} 
Defining $\alpha=A/g'(0)$, this differential equation is solved by
\begin{equation}
g(E_i)=A\,e^{\alpha\,E_i}.
\end{equation}
Using Eq.~(\ref{eq:determine}), we can easily deduce $f$ to be
\begin{equation}
f(E_i)=B\,e^{\alpha\,E_i}.
\end{equation}
Thus, we have proven the transformation given by Eq.~(\ref{eq:transformation})
to be the most general way to construct all solutions of 
Eq.~(\ref{eq:normansatz}) from one arbitrary solution.

\clearpage

\begin{table}[htbp]\centering
\begin{tabular}{c|c|c|c}
iteration&$P$ (\%)&number of iterations&max.\ variation\\\hline\hline
1--5&20&5&$1.2^5\approx 2.49$\\
6--12&10&7&$1.1^7\approx 1.95$\\
13--21&5&9&$1.05^9\approx 1.55$\\
22--30&2.5&9&$1.025^9\approx 1.25$\\
31--50&1&20&$1.01^{20}\approx 1.22$\\\hline
\multicolumn{2}{c}{}&$\sum=50$&$\prod\approx 11.46$\\
\end{tabular}
\caption{$P$ as a function of the number of iterations. We have in the actual 
computer code implemented 50 iterations where $P$ is decreasing gradually from
20\% to 1\%. The maximum variation from the $0^{\mathrm{th}}$ order estimate 
any point can get is approximately a factor 11.}
\label{tab:decrease}
\end{table}

\end{document}